\documentclass[12pt]{spieman}  % 12pt font required by SPIE;
\usepackage{amsmath,amsfonts,amssymb}
\usepackage{graphicx}
\usepackage{setspace}
\usepackage{tocloft}
\usepackage{subcaption}
\usepackage{sidecap}
\sidecaptionvpos{figure}{t}
\usepackage[symbol]{footmisc}
\usepackage{lineno}
\usepackage{soul}

\usepackage{multirow}
\usepackage{makecell}
\usepackage{{booktabs}}
\usepackage[scientific-notation=true]{siunitx}
\usepackage{collcell}
\usepackage{caption} 
\title{Learnable real-time inference of molecular composition from diffuse spectroscopy of brain tissue}

\author[a,$\dagger$,*]{Ivan Ezhov}
\author[a,$\dagger$]{Kevin Scibilia}
\author[b,c]{Luca Giannoni}
\author[f]{Florian Kofler}
\author[a]{Ivan Iliash}
\author[a]{Felix Hsieh}
\author[a]{Suprosanna Shit}
\author[e]{Charly Caredda}
\author[d]{Frédéric Lange}
\author[e]{Bruno Montcel}
\author[d]{Ilias Tachtsidis}
\author[a,g]{Daniel Rueckert}

\affil[a]{Department of Computer Science, Technical University of Munich, Germany}
\affil[b]{Department of Physics and Astronomy, University of Florence, Italy}
\affil[c]{European Laboratory for Non-Linear Spectroscopy, Italy}
\affil[d]{Department of Medical Physics and Biomedical Engineering, University College London, United Kingdom}
\affil[e]{Univ Lyon, INSA-Lyon, Université Claude Bernard Lyon 1, UJM-Saint Etienne, CNRS, Inserm, CREATIS UMR 5220, France}
\affil[f]{Helmholtz AI, Helmholtz Munich, Germany}
\affil[g]{Department of Computing, Imperial College London, United Kingdom}

\cftpagenumbersoff{figure}
\cftpagenumbersoff{table} 
\begin{document} 
\maketitle

%\begin{abstract}
\noindent\textbf{Abstract}

\noindent\textbf{Significance:}
Diffuse optical modalities such as broadband near-infrared spectroscopy (bNIRS) and hyperspectral imaging (HSI) represent a promising alternative for low-cost, non-invasive, and fast monitoring of living tissue. Particularly, the possibility of extracting the molecular composition of the tissue from the optical spectra deems the spectroscopy techniques as a unique diagnostic tool.

\noindent\textbf{Aim:}
No established method exists to streamline the inference of the biochemical composition from the optical spectrum for real-time applications such as surgical monitoring. Here, we analyse a machine learning technique for inference of changes in the molecular composition of brain tissue.

\noindent\textbf{Approach:}
We propose modifications to the existing learnable methodology based on the Beer-Lambert law. We evaluate the method's applicability to linear and non-linear formulations of this physical law. The approach is tested on data obtained from the bNIRS- and HSI-based monitoring of brain tissue.

\noindent\textbf{Results:}
The results demonstrate that the proposed method enables real-time molecular composition inference while maintaining the accuracy of traditional methods. Preliminary findings show that Beer-Lambert law-based spectral unmixing allows to contrast brain anatomy semantics such as the vessel tree and tumor area.

\noindent\textbf{Conclusion:}
We present a data-driven technique for inferring molecular composition change from diffuse spectroscopy of brain tissue, potentially enabling intra-operative monitoring.

%Diffuse optical modalities such as broadband near-infrared spectroscopy (bNIRS) and hyperspectral imaging (HSI) represent a promising alternative for low-cost, non-invasive, and fast monitoring of functional and structural properties of living tissue. Particularly, the possibility of extracting the molecular composition of the tissue from the optical spectra in real-time deems the spectroscopy techniques as a unique diagnostic tool. However, no established method exists to streamline the inference of the biochemical composition from the optical spectrum for real-time applications such as surgical monitoring.
%In this paper, we analyse a machine learning technique for fast and accurate inference of changes in the molecular composition of brain tissue. We reconsider and propose modifications to the existing learnable methodology based on the Beer-Lambert law, which analytically connects the spectra with concentrations. Moreover, we evaluate the method's applicability to linear and non-linear formulations of the Beer-Lambert law. The approach is tested on real data obtained from the bNIRS- and HSI-based optical monitoring of brain tissue. The results demonstrate that the proposed method enables real-time molecular composition inference while maintaining the accuracy of traditional linear and non-linear optimization solvers.
%\end{abstract}

% Include a list of up to six keywords after the abstract

% Include email contact information for corresponding author
{\noindent \footnotesize\textbf{*} Corresponding author,  \linkable{ivan.ezhov@tum.de} }\\
{\noindent \footnotesize\textbf{$\dagger$} The authors contributed equally}

\begin{spacing}{2}   % use double spacing for rest of manuscript

%%%%%%%%%%%%%%%%%%%%%%%%%%  body  %%%%%%%%%%%%%%%%%%%%%%%%%%
\section{Introduction}

Various biomedical applications such as histopathology or neurosurgery require access to rapid monitoring of intrinsic tissue properties. In particular, neuronavigation would benefit by having structural and functional information of brain tissue in real-time \cite{ganslandt2002neuronavigation,orringer2012neuronavigation,gerard2021brain}. Spatially-resolved maps of the tissue characteristics would allow bypassing invasive disease diagnostics, e.g., biopsy, which halt the operation. Instead, a surgical decision could be made during the operation, reducing its time and preserving a healthy brain.

Diffuse optical modalities, such as near-infrared spectroscopy and hyperspectral imaging, emerge as promising technologies to address these clinical needs \cite{sen2016clinical,giannoni2018hyperspectral}. These techniques can probe biological matter utilizing non-ionizing electromagnetic radiation within the visible and near-infrared ranges. The spectral instrumentation is inexpensive (compared to other modalities such as MRI), allows for continuous tissue monitoring, and can be easily adapted for the complex context of the operating room \cite{viderman2021near}.  

The overarching principle behind spectroscopy-based molecular characterization is to relate the reflection spectrum obtained upon illumination of the tissue surface with its optical properties. The molecules constituting the tissue have unique absorption dependency on the optical radiation frequency, and thus, the reflection should exhibit molecular absorption signatures in its frequency dependency. 

However, several other physical factors contribute to shaping the measured reflectance spectra. These can include light scattering on the surface and within tissue volume, autofluorescence, tissue inhomogeneity, and background illumination \cite{lu2014medical}. Disentangling these phenomena from a reflectance spectrum is often an ambiguous, ill-posed problem, yet it is crucial for deducing the relation between the reflection and molecular composition. Another complication is a scarcity of available studies in which optical monitoring and quantitative biochemical composition analysis are performed simultaneously. 

Analytical and statistical approaches exist aiming to unmix optical spectra into the physical phenomena defining the spectra profile under a limited data regime \cite{tzoumas2017spectral,dobigeon2016linear,lu2015estimation,bateson1996method,ezhov2023identifying,yokoyama2003estimation,gerstner2012hyperspectral}. A large number of methods mitigate the data scarcity by introducing a physical prior to establish the spectrochemical link. Typically, the modified Beer-Lambert law \cite{stratonnikov2001evaluation,oshina2021beer} is used to provide such a link by describing the incoming light's energy dissipation as an exponentially decaying function:

\begin{equation}
\log (I_R(\lambda) / I_0(\lambda)) = - \left[ \sum_i{c_i\mu^i_a(\lambda)}+s\mu_s(\lambda) \right]l  + U
\end{equation}

\noindent Here, $I_R(\lambda)$ and $I_0(\lambda)$ are the intensities of the reflected and the incoming light, $\mu_a$ and $\mu_s$ are the absorption and scattering coefficients, the index $i$ denotes the molecule constituting the tissue such as water, fat, hemoglobin and cytochromes, $c_i$ denotes the corresponding concentration (e.g. as volume fraction), and $s$ is the weight of scattering in the total light energy dissipation. The remaining quantities are $\lambda$, which is the light wavelength, $l$ is the light pathlength\footnote{Even though several works demonstrate the importance of wavelength-dependent definition of the pathlength \cite{bahl2023comparative}, in what follows, for simplicity, we opted for constant pathlength independent of the wavelength.}, and $U$ describes other physical factors contributing to the energy dissipation of the incoming light or other sources of the optical signal captured by the light detectors or cameras. 

Typical molecules whose changes in concentration are inferred include oxyhemoglobin, deoxyhemoglobin, and cytochrome-c-oxidase (CCO)\cite{giannoni2018hyperspectral}. Measurement of the former two chromophores can reveal e.g. oxygenation status of the brain, and can help determine hypoxic or hyperoxic conditions. CCO is a fundamental metabolic molecule correlated to ATP production during cellular respiration, which has previously complemented obtained hemodynamic information in various applications\cite{giannoni2018hyperspectral, bale2016jobsis}.

Now, in the case of changes in molecular composition over the course of optical monitoring, one can assume that the effects contributing to $U$ either stay constant (e.g., which is a fair assumption for ambient illumination) or change notably less than the total absorption\footnote[4]{In certain scenarios, even changes in scattering are not expected since it is a rather bulk effect dependent on the density of the probed matter rather than molecule-specific one.}. % For brain tissue, the total concentrations of hemoglobin and especially cytochrome stay nearly constant that, in turn, should not affect the tissue density.}. 
Under this assumption, subtraction of two reflection spectra, $\log I_R^2 - \log I_R^1$ (traditionally, these have been two different points in time, but this could also be true for different points in space), would cancel out or make negligible the term $\delta U=U^2-U^1$ in the following equation:
\begin{equation}
\log (I_R^2(\lambda) / I_R^1(\lambda)) = - \left[ \sum_i{\delta c_i\mu^i_a(\lambda)}+\delta (s\mu_s(\lambda)) \right]l + \delta U \ %\sum_i{\delta c_i\mu^i_a(\lambda)}>>\delta U
\label{dBBL}
\end{equation}

\noindent In such a differential form, the modified Beer-Lambert law can now be used to identify molecular composition. For this, standard least-squares optimization algorithms (or non-negative matrix factorization \cite{lu2015estimation,caredda2024priori}) can be employed to minimize the difference between the real spectra and the spectra obtained from the modified Beer-Lambert law. As a result of the minimization, the optimal values of the set of concentration changes $\{\delta c_i=c_i^2-c_i^1\}$ are obtained (alongside the scattering parameters). 

%Another family of methods is based on statistical techniques such as principal component analysis (PCA) \cite{bateson1996method,yokoyama2003estimation,gerstner2012hyperspectral,ezhov2023identifying}. Here, by analyzing a large dataset of reflectance spectra using the PCA, one identifies the importance of each wavelength to explain the variance in the dataset. Assuming that from the physics point of view, the variance in the reflectance spectra is mainly due to varying molecular absorption, one can try to correlate the output of the PCA analysis with the absorption spectra. In \cite{ezhov2023identifying}, the authors discuss how such correlation can reveal the relative difference in concentration of chromophores across different tissue types.

The overarching drawback of this approach is the computational time it takes to infer the biochemical composition. For example, the optimization methods take a subsecond time to infer the composition of a single spectrum containing a number of wavelengths typical for bNIRS and HSI (a few hundred). However, for real-time applications particularly in the case of HSI modality, one needs to solve the optimization task in a subsecond time for as many spectra as there are spatial pixels, as every pixel contains its own spectrum. The number of pixels on a hyperspectral image can be easily in the order of $10^5$-$10^6$. Providing subsecond timings for simultaneous inference on such an amount of spectra poses a challenge for traditional methods.  

\paragraph{Contributions.}

There are numerous studies analyzing the application of machine learning methods to achieve fast inversion of the physical models based on Beer-Lambert law or Monte-Carlo simulations \cite{chen2015efficient,wirkert2016robust,panigrahi2019machine,fredriksson2020machine,hokr2021machine,manojlovic2023rapid,scarbrough2024designing}. Predominantly they imply training a machine learning model on synthetic data generated by following the chosen physical formalism, and then evaluating the trained model on real spectra. While proven to work for the use cases mentioned in the cited works, this approach might be inferior as synthetic data generators likely underestimate the complexity of real data. To mitigate this, we tested different strategies for model training using only the \textit{synthetic data} as in previous works or incorporating \textit{real data} via traditional optimization in the training procedure. 

Second, we test the proposed method on its ability to approximate physical models of varying complexity: \textit{linear} (absorption only) and \textit{non-linear} (absorption combined with scattering). While evaluated independently in previous works, here we also analyse our approach to explicitly compare both models in terms of spectral fit. This comparison is motivated by a desire to elucidate the conditions under which the linear model (that can be easily solved using, e.g. pseudoinverse) is appropriate for describing the light-brain matter interaction process, and where inclusion of the scattering is necessary.

%We test two types of deep-learning-based approaches which differ in the way the training dataset is collected: a) directly by generating synthetic spectrum-concentrations pairs from the modified Beer-lambert law or b) by inferring the molecular composition from real spectra by means of the traditional least-squares optimization and then use the pairs of real spectra and inferred concentrations for a neural network training.

Third, given that the work is carried out within the HyperProbe project \cite{giannoni2023hyperprobe} aiming to achieve real-time brain tissue monitoring, the present paper evaluates the \textit{computational timing} for the biochemical composition inference across different methods and hardware platforms. To reinforce the comparison, in contrast to previous works manually choosing the hyperparameters' values of the machine learning methods, we used the AutoML technique \cite{liaw2018tune} to identify the most optimal hyperparameters set.

Finally, to our knowledge, this is the first work that applies a neural-network-based approach to provide real-time inference of chromophore composition from \textit{in-vivo brain tissue} spectroscopy measurements \cite{mamouei2021empirical,chou2021deep,yap2022adaptive,qureshi2016comparison,altmann2011nonlinear,karamavucs2019newborn,bian2020multiple,ewerlof2022multispectral,zhou2022real}. We evaluate and discuss applicability of the method on broadband NIRS (transmission mode) \cite{kaynezhad2019quantification} and hyperspectral (reflection mode) \cite{fabelo2019vivo} measurements of brain tissue.

\section{Method}

As mentioned in the introduction, inference of absolute chromophore concentrations from an optical spectrum is a challenging task due to multiple physical effects shaping the reflection spectrum. Thus, we instead aim to predict the changes in concentrations from changes in the spectra, Eq. \ref{dBBL}. 

\begin{figure}[h]
\centering\includegraphics[width=\textwidth]{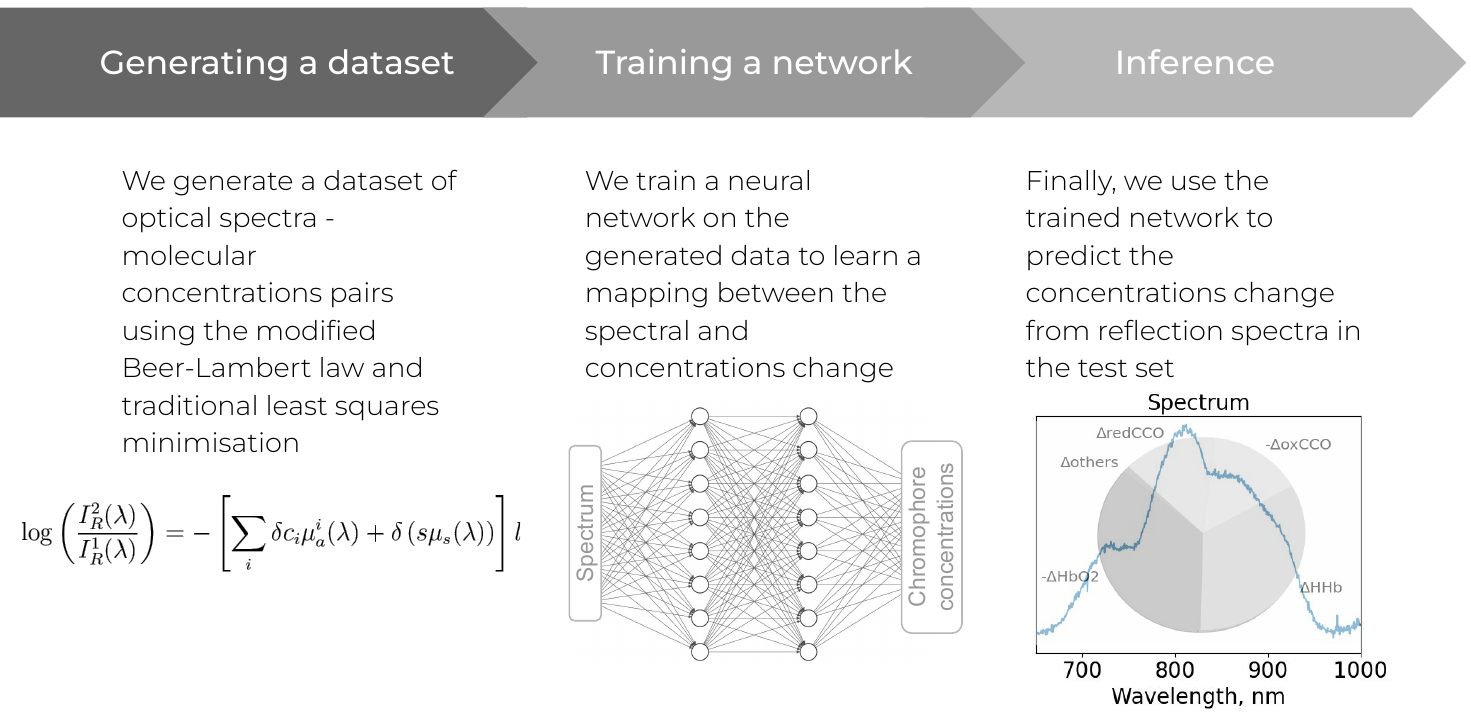}
\caption{The general pipeline describing the learnable approach for inferring concentrations' changes of molecules such as reduced and oxidized cytochrome-c-oxidase, oxy- and deoxy-hemoglobin $\{\delta c_i\}=\{ \delta redCCO, \delta oxCCO, \delta HHb, \delta HbO_2, etc \}$. The pipeline involves training on a dataset which is generated by the means of a modified Beer-Lambert model. According to the model, the light reflection $I_R$ is shaped by the absorption $\mu_a$ and scattering $\mu_s$ phenomena.}
\end{figure}

\subsection{Datasets creation.}

In our method, Fig. 1, we use a supervised data-driven approach by creating a dataset of attenuation-concentration pairs to train a neural network (by attenuation, or more precisely the change of it, we imply the logarithm of the reflection: $\Delta A = \log I_R^2 - \log I_R^1$). We employ two different strategies to create the dataset:

a) The first strategy directly utilizes the modified Beer-Lambert law to generate the training dataset pairs $(\Delta A, \{ \delta c_i\})$ with $\Delta A$ being the difference in attenuation between two spectra and $\{\delta c_i\}$ the corresponding differences in concentration of chromophores, Fig. 2 (a). For each chromophore, we randomly sample values for changes in the molecular composition $\{ \delta c_i\}$ using the uniform distribution within physiologically plausible ranges. These ranges were determined based on values typically used in the literature, and further details are provided in the appendix.
 
If scattering is included in Eq. \ref{dBBL}, we may assume it to be of rational form: 
\begin{equation}
s\mu_s(\lambda) = s\left( \frac{\lambda}{500nm} \right)^{-b}
\label{scatter}
\end{equation}
with the scaling of the anisotropy $g=0.9$ included in $s=s'/(1-g)$ \cite{jacques2013optical, giannoni2018hyperspectral}. In the differential form, we obtain 
\begin{equation}
\delta (s\mu_s(\lambda)) = s_2 \left( \frac{\lambda}{500nm} \right)^{-b_2} - s_1 \left( \frac{\lambda}{500nm} \right)^{-b_1}
\end{equation}

and therefore also uniformly sample parameters $s_1, s_2, b_1, b_2$ within plausible ranges \cite{jacques2013optical}. Subsequently, we input the obtained values into the modified Beer-Lambert law to obtain synthetic differential attenuations $\Delta A(\lambda)$. This difference in attenuation $\Delta A$ as input and the corresponding $\{\delta c_i\}$ as output are then used for training.

\begin{figure}[t]
    \centering

    \includegraphics[width=\textwidth]{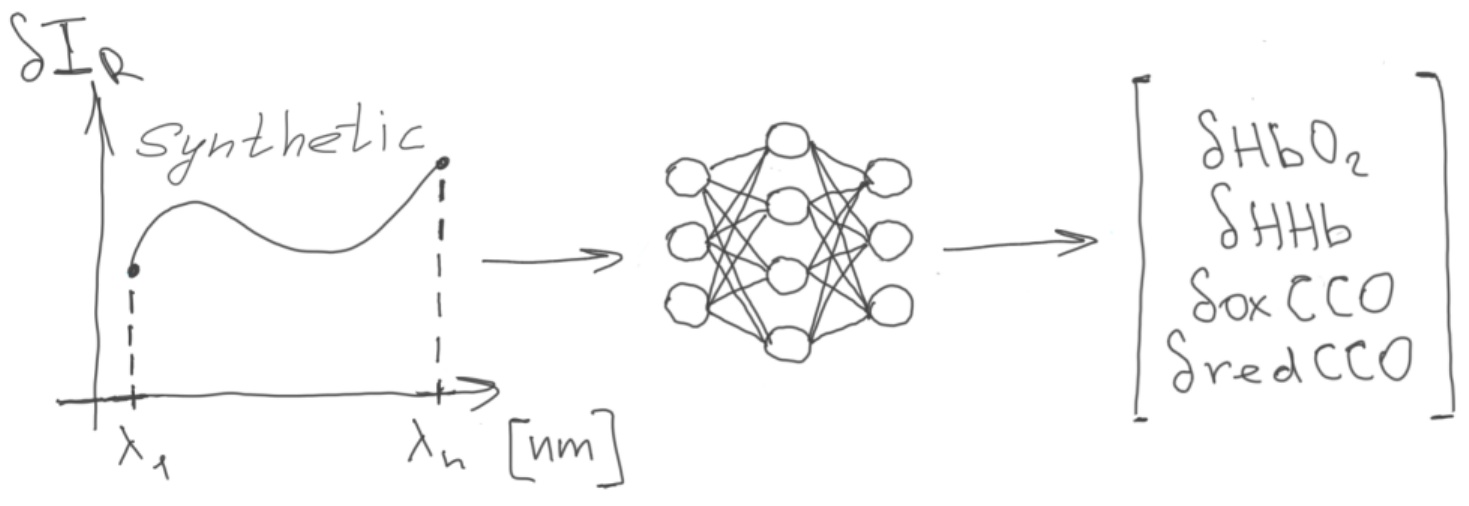}
    \includegraphics[width=\textwidth]{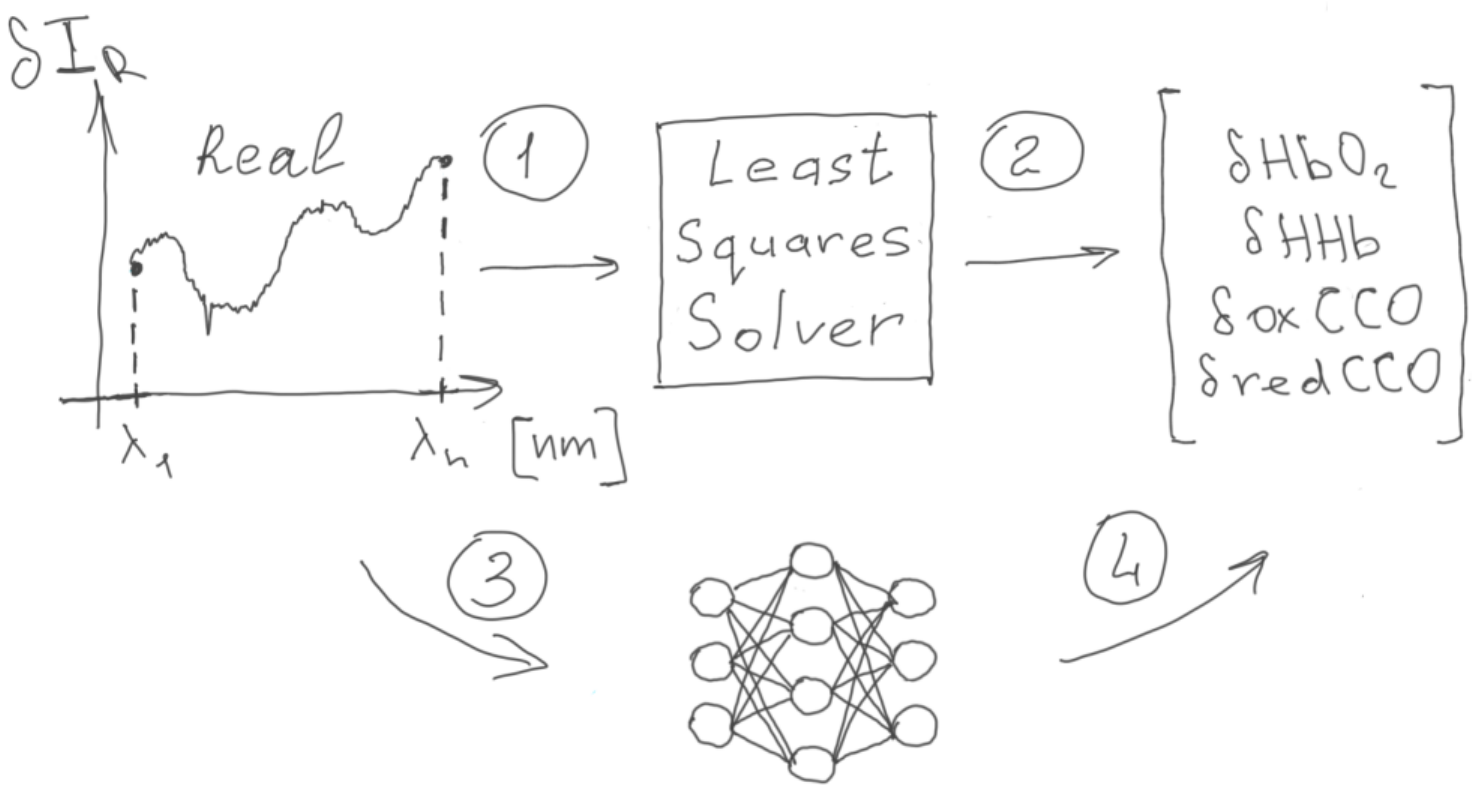}

    \caption{Two strategies for collecting the training dataset. Strategy (a) in which we train a network on synthetic attenuation-concentration pairs generated from the modified Beer-Lambert law. Strategy (b) in which the training is performed on pairs of real attenuation and concentrations obtained through the least-squares fit to the corresponding real spectra.}
\end{figure}

b) Given that the distribution of the synthetic spectra obtained according to the strategy described above can be notably different from the distribution of real spectra, this can result in an unsatisfactory network prediction accuracy. Therefore, in addition to (a), we evaluate another strategy for creating a dataset trying to bridge the gap between the physical model and real data, Fig. 2 (b). For this, we use traditional least squares minimization to fit the changes in the real reflectance spectrum with the modified Beer-Lambert law. The concentrations $\{\delta c_i\}$ found upon the optimization and the corresponding $\Delta A$ constitute the training samples.

\subsection{Network and optimization details.} The training was performed using a multi-layer perceptron (MLP \cite{rumelhart1986learning}) neural network for both approaches. The network takes as input a one-dimensional vector of attenuation difference and outputs molecular concentration changes.
 
We trained both networks with early stopping when they reached convergence. To find the optimal network architecture, we used the Ray Tune library \cite{liaw2018tune} to validate different MLP architectures (width, number of hidden layers, and activation functions), learning rates, and batch sizes. More details regarding the networks and the training procedure can be found in the appendix.

Least squares optimization for the Beer-Lambert law, excluding the scattering effect, can be performed via multiplication of the observed attenuation with the pseudoinverse of the absorption coefficients \cite{giannoni2018hyperspectral}. In order to perform the nonlinear least-squared optimization for the Beer-Lambert law model including scattering, we used the publicly available solver of the SciPy library \cite{virtanen2020scipy}. 
We used the least-squares minimization obtained predictions as the ground truth to validate all the trained networks.

% \begin{equation}
% A = \begin{bmatrix}
% \mu_1(\lambda_1) & \mu_1(\lambda_2) & \cdots & \mu_1(\lambda_m) \\
% \mu_2(\lambda_1) & \mu_2(\lambda_2) & \cdots & \mu_2(\lambda_m) \\
% \vdots & \vdots & \ddots & \vdots \\
% \mu_n(\lambda_1) & \mu_n(\lambda_2) & \cdots & \mu_n(\lambda_m)
% \end{bmatrix}
% \end{equation}

% \begin{equation}
% X = \begin{bmatrix}
% \delta c_1 \\
% \delta c_2 \\
% \vdots \\
% \delta c_n
% \end{bmatrix}
% \end{equation}

% \begin{equation}
% B = \log(I_R^1) - \log(I_R^2)
% \end{equation}

% The linear system of equations is given by:
% \begin{equation}
% AX = B
% \end{equation}

\subsection{Data.} For our experiments, we applied two types of Beer-Lambert law formulation, with and without scattering, to two types of spectral datasets: broadband NIRS data for which the spectra were measured in light transmission mode \cite{kaynezhad2019quantification} and hyperspectral data which were obtained in non-contact reflection mode \cite{fabelo2019vivo}.

\begin{figure}[b!]
\centering\includegraphics[width=\linewidth]{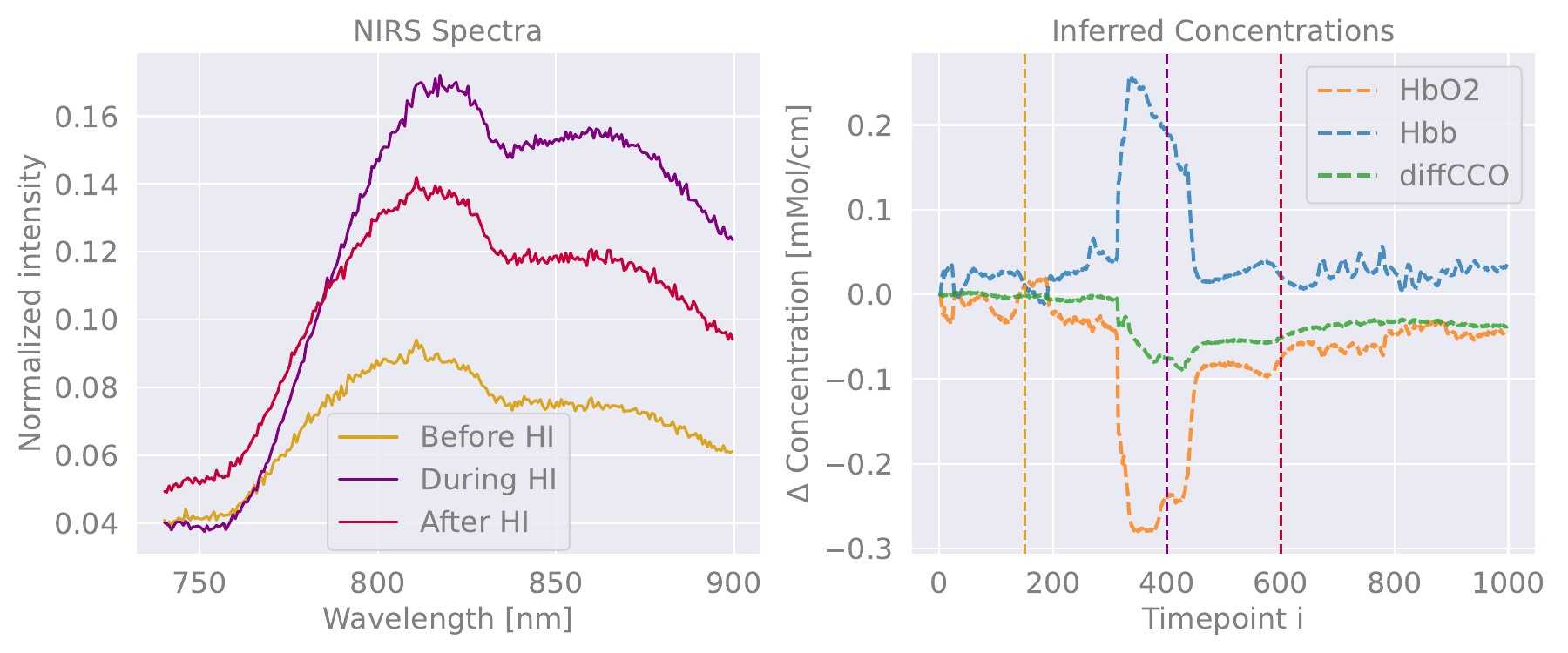}
\caption{Optical spectra from the broadband bNIRS study in \cite{kaynezhad2019quantification} before, during and after inducing HI in the piglet's brain (left). On the right, predictions of the molecular concentration change $\{ \delta c_{HbO_2},  \delta c_{HHb}, \delta c_{\text{diffCCO}}\}$ over the course of the optical monitoring (for the Beer-Lamber model \textit{without} scattering). The vertical lines denote the time points corresponding to the normalized reflection spectra on the left.}
\label{ilias_spect}
\end{figure}

\subsection{Broadband NIRS.} The first dataset is composed of broadband NIRS spectra from a study analyzing 27 piglets' brains in which a hypoxia-ischemia (HI) state was induced \cite{kaynezhad2019quantification}. The piglets were monitored for several hours, during which the carotid arteries were surgically isolated, and a stepwise hypoxia took place for 15-20min. This produced a significant hypoxic-ischaemic effect that changed the metabolic status of the brain and, in some instances, caused further brain injury. The details of the intervention protocol are described in \cite{kaynezhad2019quantification}. 
%The measurements contain around eight thousand spectra per piglet, of which we use the first thousand measurements (corresponding to the first $\sim$2.5 hours of measurement) where HI and recovery after HI are visible. 
The optical device used in the study utilizes a miniature light source and a customized high-throughput miniature spectrometer, connected to high numerical aperture optical fibers. The measurements contain around eight thousand spectra per piglet. The distance between each measured time point is between 10.0 and 10.5 seconds. We use the first thousand measurements, i.e. we only consider the first $\sim$2.5 hours of measurement. For all piglets, this is sufficient to observe HI and recovery after HI.
As Eq. \ref{dBBL} requires defining a baseline spectrum, analogously to \cite{kaynezhad2019quantification}, we used a spectrum at the very beginning of optical monitoring (i.e., before HI) for the baseline. We normalized the spectra with respect to dark noise and white reference. The normalized bNIRS spectra before and after the intervention inducing hypoxia are shown in Fig. \ref{ilias_spect} (left), and predictions of the concentrations change over the course of the optical monitoring are shown in Fig. \ref{ilias_spect} (right). 
Out of the 27 piglets in the dataset, 25 had data available during HI, such that 19 were used for training, two for validation, and four for testing.
For this dataset, we predict three types of molecules: oxyhemoglobin, deoxyhemoglobin, and differential cytochrome-c-oxidase (CCO)\footnote[2]{As the total CCO concentration may be assumed to not change within a few hours, the oxidized-reduced difference spectrum $\mu_a^{\text{diffCCO}}$ may be used to infer changes of both oxidized and reduced CCO \cite{bale2016jobsis}. }, i.e. $\{\delta c_i \} = \{ \delta c_{HbO_2},  \delta c_{HHb}, \delta c_{\text{diffCCO}}\}$, where $\delta c_{\text{diffCCO}}=\delta c_{oxCCO} - \delta c_{redCCO}$. We neglected the potential contribution to the spectra from water and fat due to their minimal change in concentrations during the 2.5 hours of monitoring \cite{kaynezhad2019quantification,giannoni2020investigation}. Note that we assume unitary pathlength in our experiments, which results in units of [mM/cm] and [1/cm] for the inferred concentrations. In the NIR range, it has been shown that the pathlength is semi-constant \cite{kaynezhad2019quantification}, which effectively leads to a simple rescaling in our concentrations when using this assumption. This can also be observed when comparing the inferred exemplary concentrations from Fig. \ref{ilias_spect} (right) with pathlength-corrected concentrations in previous work \cite{kaynezhad2019quantification}.

\subsection{Hyperspectral data.}
\begin{figure}[t!]
\centering
\includegraphics[width=\linewidth]{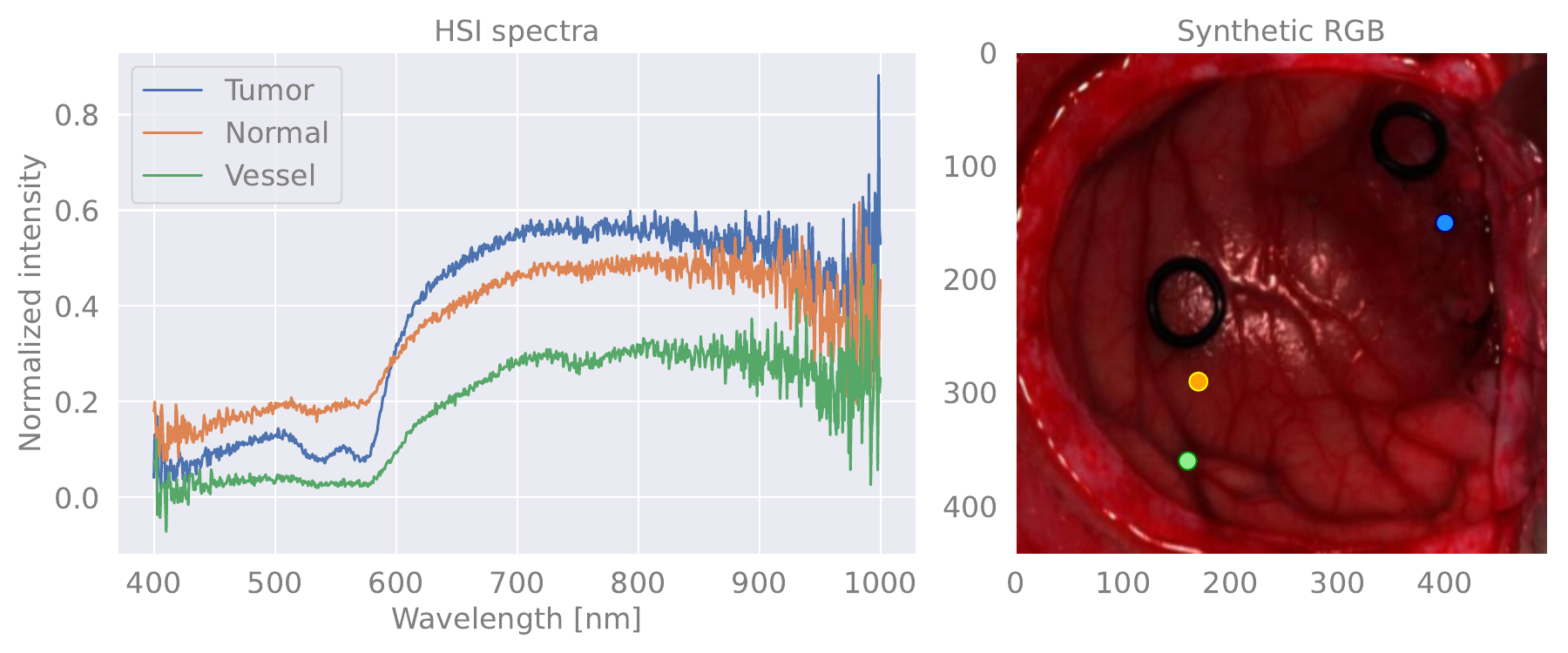}
\caption{Optical spectra from the HSI study of patients diagnosed with glioma \cite{fabelo2019vivo} 
for different tissue types: tumor, normal tissue, and blood vessels (left). On the right is a typical RGB image of the brain surface, which is obtained from the HSI volume. The dots correspond to the spectra on the left image. The black circles on the RGB image are rubber rings that surgeons used to mark tumomr and healthy tissues.}
\label{helic_spect}
\end{figure}

The second dataset we used consists of hyperspectral data from the Helicoid project \cite{fabelo2019vivo}. The Helicoid dataset comprises brain HSI images obtained in surgical conditions from 22 patients diagnosed with glioma. The optical instrumentation is based on pushbroom technique and a silicon CCD detector array as a camera. The HSI images provide a high spectral resolution of 826 bands spread between 400 and 1000 nm and a 2D spatial resolution of a few hundred pixels in each dimension. The images were also expert-annotated into three tissue classes: normal and tumor tissues, as well as blood vessels.
Typical hyperspectral image and corresponding spectra are shown in Fig. \ref{helic_spect}. Different from the bNIRS dataset, we used a spectrum of the pixel belonging to the blood vessel class as a baseline spectrum\footnote[4]{The blood vessel was used as a reference for a couple of reasons.
The blood vessel is clearly distinguishable from the other two tissue types, tumor and non-tumor tissue. These two types of tissue are highly heterogeneous, e.g. within a pixel area, they can have small capillaries, leakage of blood, agglomeration of dead cells, etc. In contrast, the blood vessel pixels, especially the ones belonging to large arteries, are less heterogeneous.
Moreover, it is assumed that blood vessels do not possess cytochrome molecules. Thus, it is a better reference when one sets a goal of detecting the presence of cytochromes in the brain matter.
}. We then subtracted the baseline spectrum from all other spectra in the same image. In other words, we performed the differential spectroscopy \textit{not in time but in space}.

\begin{figure}[t!]
\centering
\includegraphics[width=\linewidth]{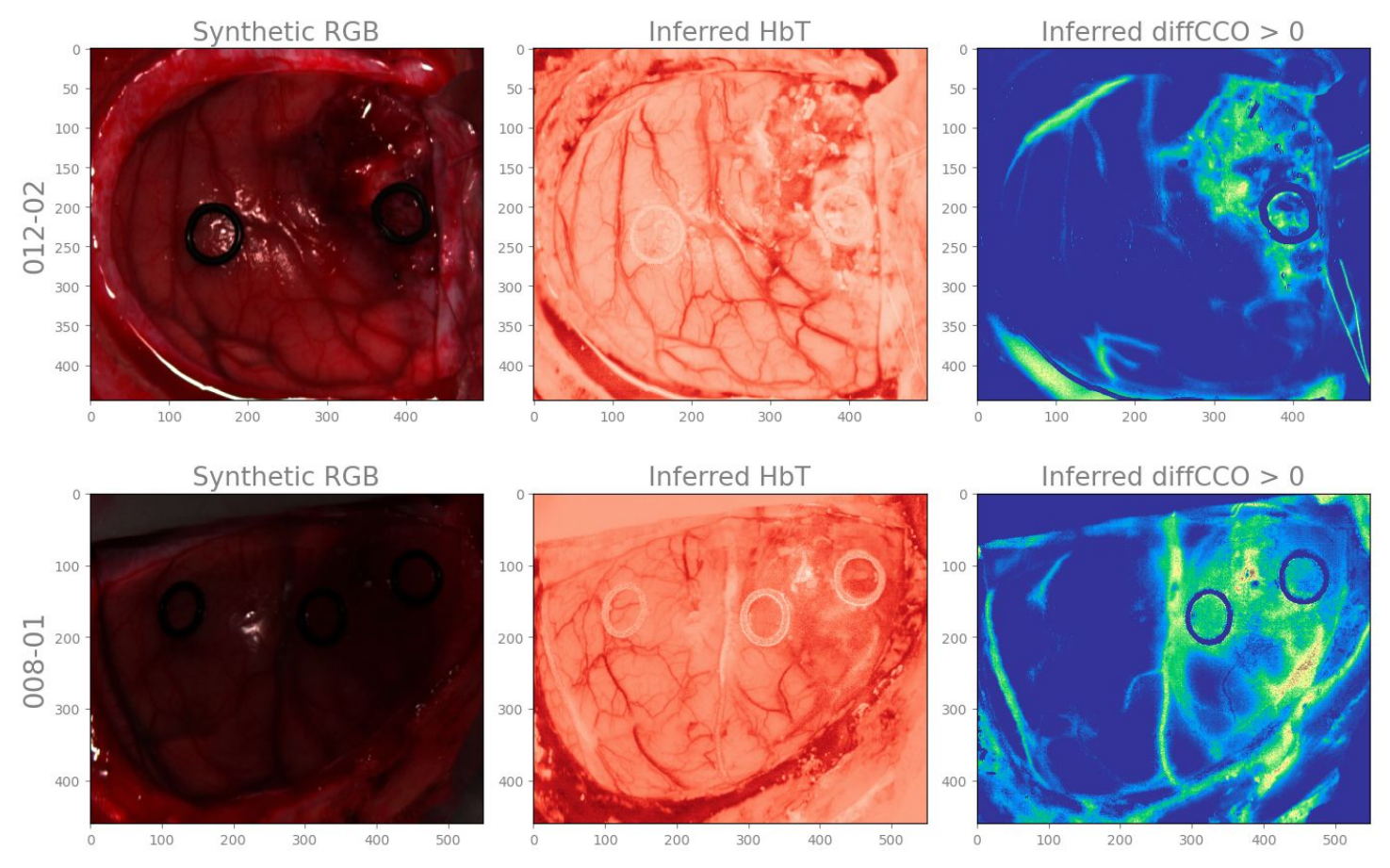}
\caption{Examples of HSI images of two patients, shown with respective patient ID, from the Helicoid dataset \cite{fabelo2019vivo} (left). Each pixel in the shown 2D image possesses a spectral signature with 826 bands. From this signature, we predict the molecular concentration change for hemodynamic $\delta c_{HbT} = \delta c_{HbO_2} + \delta c_{HHb}$ (middle), and metabolic $\delta c_{\text{diffCCO}}=\delta c_{oxCCO} - \delta c_{redCCO}$ characterization (right). Here, we use the Beer-Lambert model \textit{with} scattering, as it provides a closer fit to real spectra than the model without scattering. We observe that performing the spectral unmixing on the HSI measurement of brain tissue allows us to better contrast the vessel tree (middle) and tumor area (right) than on the RGB image.
}
\label{helic_res}
\end{figure}

Besides predicting oxyhemoglobin $\delta c_{HbO_2}$ and deoxyhemoglobin $\delta c_{HHb}$, we again infer the differential cytochrome-c-oxidase concentration due to its role in capturing oxidative metabolic activity. We separately predicted oxidized cytochrome-c-oxidase and reduced cytochrome-c-oxidase, as the total CCO concentration may not be assumed to remain constant in space. We also predict water and fat since, for these molecules, one cannot assume minimal concentration change across different tissue types as in the case of the bNIRS spectra. For reference, the absorption spectra can be found in the appendix. 

Fig. \ref{helic_res} showcases examples of molecular inference for the HSI images from the Helicoid dataset.
%Thus, in addition to oxyhemoglobin, deoxyhemoglobin, and differential cytochrome-c-oxidase, we also predict water and fat $\{\delta c_i \} = \{ \delta c_{HbO_2},  \delta c_{HHb}, \delta c_{\text{diffCCO}},  \delta c_{Water}, \delta c_{Fat}\}$, since for these molecules one cannot assume minimal concentration change across different tissue types as in the case of the NIRS spectra. Fig. \ref{helic_res} showcases examples of molecular inference for the HSI images from the Helicoid dataset.
%Out of the XXX in the dataset XXX were used for training, XXX for validation, and XXX for testing.
Out of the nine patients with glioblastoma in the dataset, six with distinct class labeling were chosen, and three patients were used for training, one for validation, and two for testing. Note that patients might have multiple images taken, and different images from the same patient were assigned to the same training, validation, and test set to avoid set contamination. Therefore, the training set consists of five, the validation set of one, and the test set of three images. 

\section{Results}
\subsection{Scattering vs Non-scattering.}

First, before discussing the learnable methods for molecular inference, we test different Beer-Lambert law formulations - with and without scattering - to elucidate limits of applicability of both models. For the case of piglets undergoing HI, it is widely assumed that the 780nm to 900nm range is predominantly dominated by absorption, with scattering being only a minor contributor to the overall measured spectrum \cite{kaynezhad2019quantification}. As measurements in the piglet dataset below the 780nm threshold were available, we opted to extend the model fitting range from 740nm to 900nm. This test is motivated by our desire to assess whether a linear model (without scattering) would still be sufficient to describe the broader spectroscopy measurement of brain tissue. 

%Also, we wanted to evaluate if networks can substitute not only linear solvers but also non-linear ones (with scattering). 

\begin{figure}[t]
%\centering\includegraphics[width=14cm]{spect_fit.png}
\centering\includegraphics[width=\textwidth]{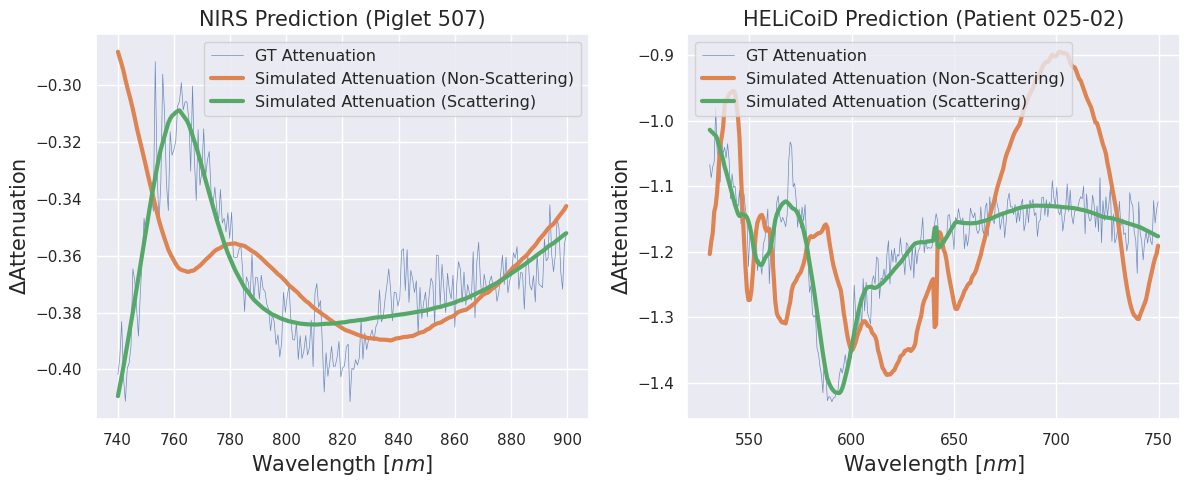}
\caption{Comparison between predictions using linear (no scattering) and non-linear (with scattering) models for bNIRS (left) and HSI (right) spectra. The ground truth (GT) attenuation is computed from the real spectra difference. The inclusion of scattering into the formulation of the Beer-Lambert law notably improves the spectral fit to real data.}
\label{spect_fit}
\end{figure}

\begin{figure}[b]
%\centering\includegraphics[width=14cm]{spect_fit.png}
\centering\includegraphics[width=\textwidth]{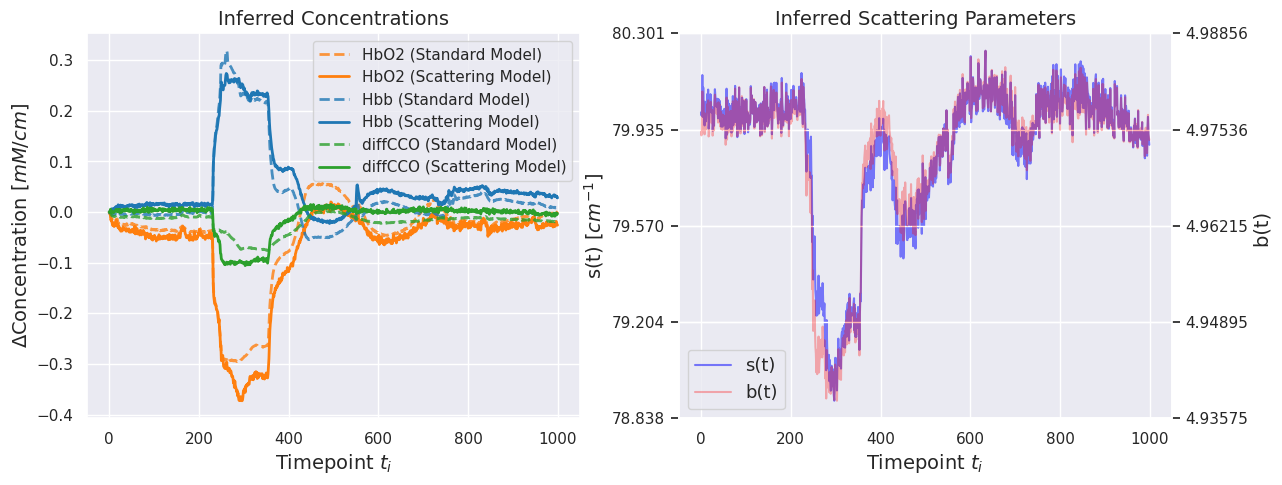}
\caption{On the left, comparison between chromophore predictions using the linear model without scattering ("Standard model") and non-linear model with scattering ("Scattering model") for the bNIRS dataset. On the right, the predicted coefficients $s$ and $b$ for the scattering term in Eq. \ref{scatter}. Note that the purple curve represents an overlap across the two red and blue curves.
}
\label{spect_fit_scat}
\end{figure}

Fig. \ref{spect_fit} and \ref{spect_fit_scat} show spectral fits and inferred molecular concentrations using both formulations. The model with scattering provides a clearly better fit. It allows us to better describe the peak around 760 nm for the bNIRS data, while for the HSI data, the inclusion of scattering is often merely necessary for an accurate fit of the spectra in this wavelength range. This finding is consistent across the dataset, as shown in table \ref{tab:mae_table}.

\begin{figure}[h]
\centering
\includegraphics[width=\textwidth]{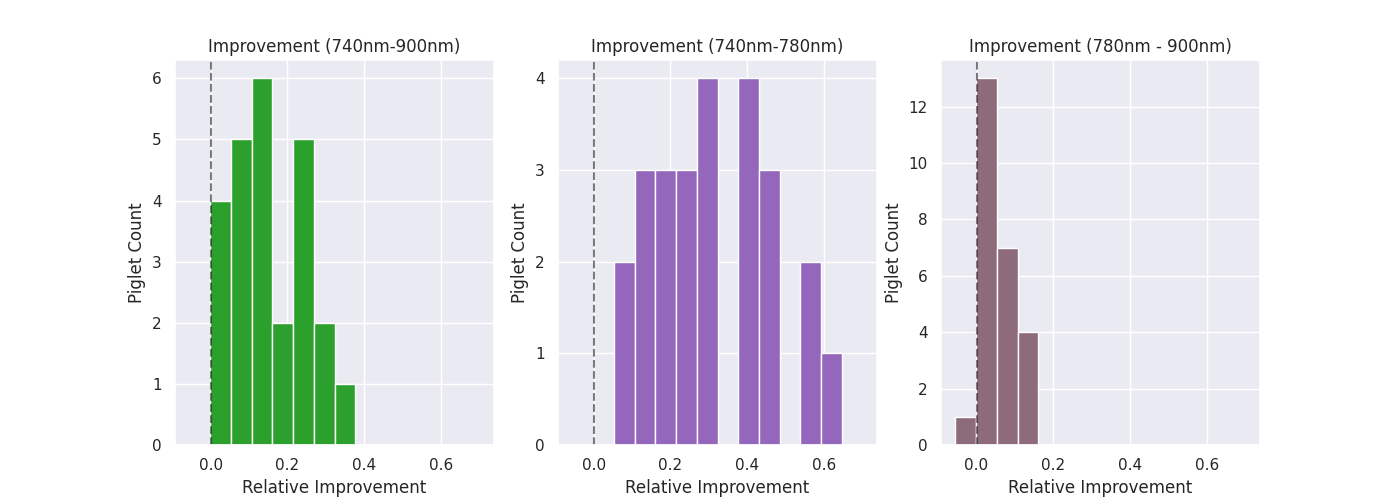}
\caption{Histograms showing mean relative improvement of the spectral fit MAE for presented scattering model, compared to the linear model, across all piglets in the dataset for all wavelengths (left), in the range 740nm-780nm (middle), and in the range 780nm-900nm (right) in the broadband NIRS dataset.
The x-axis represents the relative improvement $r$ between the two models, and the y-axis shows the number of piglets that achieved the corresponding mean relative improvement $r$. The dashed line signifies improvements below zero, i.e. cases where the spectral fit worsened.}
\label{relative}
\end{figure}

To show that the model with scattering can significantly improve model fits, especially for the higher frequency portion of the spectrum, we evaluate the relative improvement $r$ in terms of the spectral fit of the scattering model compared to the non-scattering model. We use mean absolute error (MAE) as a measure of the fit and compute it for all piglets across different spectrum bands.

The spectral fit MAE is calculated by
\begin{equation}
    \text{MAE} = \frac{ \sum_{i=1}^{n} | \Delta A_{model}(\lambda_i) - \Delta A_{data}(\lambda_i) |}{n}
\end{equation} where $\Delta A_{model}(\lambda_i)$ represents the model-inferred attenuation at wavelength $\lambda_i$, and $\Delta A_{data}$ the real measured attenuation, respectively. The relative mean improvement is computed by $$r = \frac{\text{MAE}_{scatter} - \text{MAE}_{linear}}{\text{MAE}_{linear}}$$ comparing the improvement in mean error between the scattering and linear models, computed across all timepoints.

%The results of such computation for all piglets are shown in figure \ref{relative}. We observed that a mean relative improvement is only worse by 0.6\%, and the spectral fit was still better for the of 15.7\% over the full fitting range. Such improvement is especially noticeable in the 740nm to 780nm range, where the mean improvement of the distribution almost doubled at 30.8\%. The spectral fit does not improve significantly in the 780nm to 900nm range, with the mean relative improvement of the distribution being at merely 5.6\%. These results were within our expectations due to the scattering contribution in the higher frequency range.
%For one of the 25 piglets, we observed that the spectral fit slightly worsened in the 780nm to 900nm range through the nonlinear model. However, the fitting MAE  overall range and in the 740nm-780nm range. We therefore can still confidently conclude that the presented model is able to fit the piglets' measured spectra more closely, especially for presumed scattering-dominated bands.

The results of such computation for all piglets are shown in figure \ref{relative}, where we observed a mean relative improvement of 15.7\% over the full fitting range. Such improvement is especially noticeable in the 740nm to 780nm range, where the mean improvement of the distribution almost doubled at 30.8\%. The spectral fit does not improve significantly in the 780nm to 900nm range, with the mean relative improvement of the distribution being at merely 5.6\%. For one of the 25 piglets, we observed that the spectral fit slightly worsened in the 780nm to 900nm range through the nonlinear model. However, the fitting MAE is only worse by 0.6\%, and the spectral fit was still better for the overall range and in the 740nm-780nm range. We therefore can still confidently conclude that the presented model is able to fit the piglets' measured spectra more closely, especially for presumed scattering-dominated bands.

The necessity of the scattering consideration in the Beer-Lambert model for the HSI data can be explained by the more pronounced contribution of the scattering process. For the HSI data, we infer the difference in molecular composition between different spatial locations on an image, i.e., between different tissue types. The scattering property across brain tissues can significantly vary, and thus, the scattering shapes markedly the differential spectra. In contrast, for the bNIRS data, we perform the differential spectroscopy analysis not in space but in time (comparing two spectra for the same location taken at different time points), meaning that the molecular inference is performed for the same tissue type.

In conclusion, we find that the non-linear model is especially helpful in describing scattering-dominated bands. However, the linear model may still be used when absorption is the prevalent physical effect.

\begin{figure}[h!]
   \includegraphics[width=\textwidth]{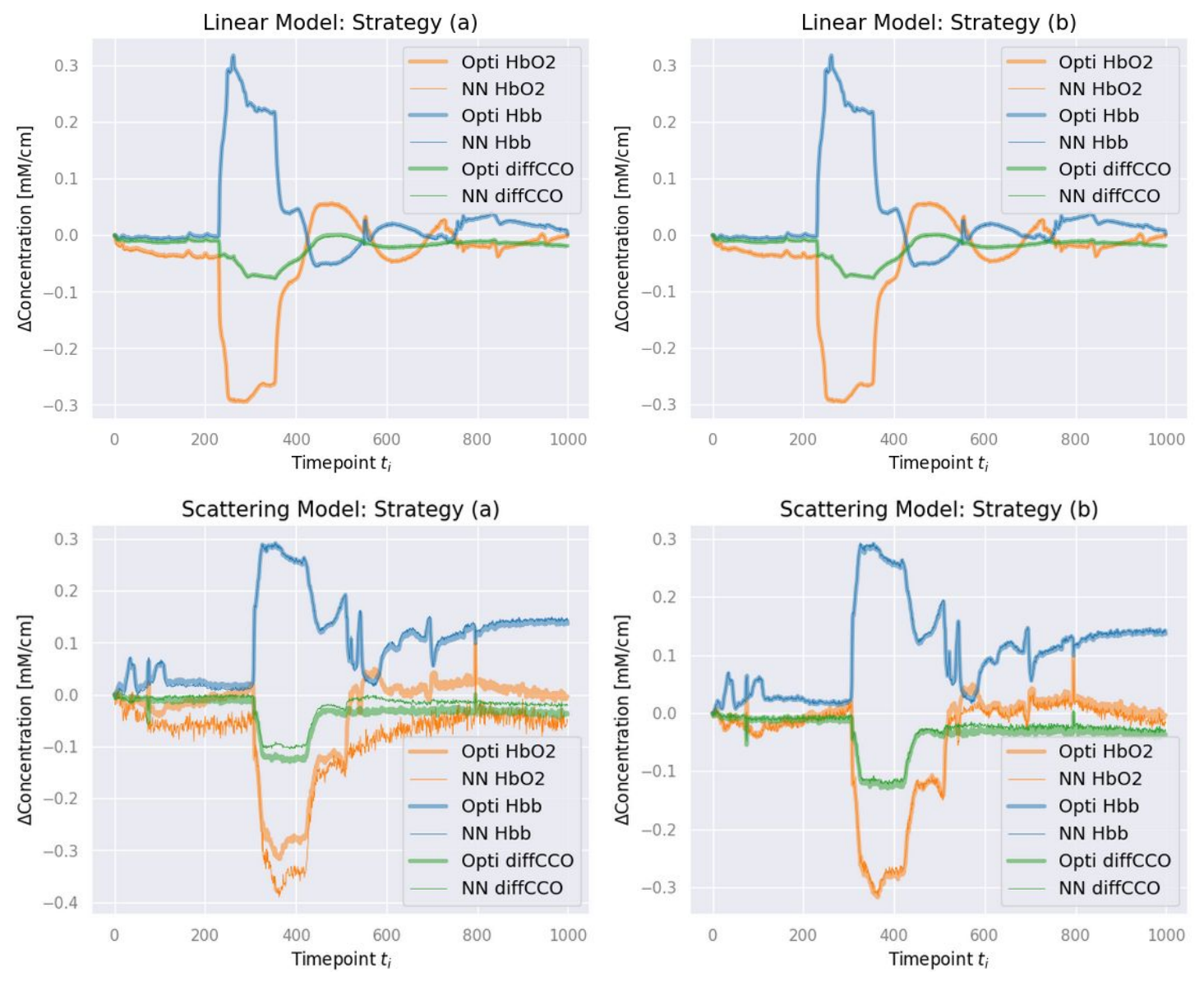}

\caption{Comparison between inference of the molecular composition using the standard optimization methods and proposed network-based inference for training strategy (a) and (b) on the bNIRS dataset. The top row compares both strategies when using the linear model, where highly accurate neural network predictions are visible in both cases.
The bottom row compares the strategies when using the non-linear model, with strategy (b) delivering noticeably more accurate predictions.
%TODO: Explain in text why strategy (b) is considerably better for the non-linear model (i.e. due to non-linearity of the problem).
%For both scenarios, the proposed strategy (b) is more accurate than strategy (a), with predictions almost indistinguishable from the standard solver.
}
\label{StrAB-NIRS}
\end{figure}

\begin{figure}[h!]
   \includegraphics[width=\textwidth]{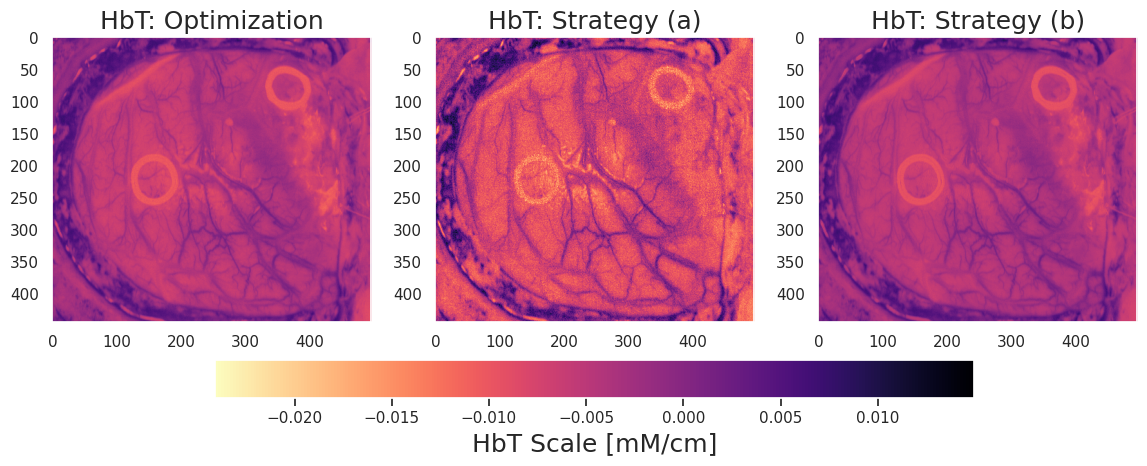}

\caption{ Comparison between inference of the molecular composition using the standard optimization methods (Opti) and proposed network-based inference (NN) for training strategy (a) and (b) on the Helicoid dataset. The top row compares strategy (a), left, with strategy (b), right, for the bNIRS dataset. The figure compares the inference of the hemodynamic signal of the optimization-based result, left, with strategy (a), middle, and strategy (b), right. Strategy (b) significantly improves results when using the non-linear model.
}
\label{StrAB-Helicoid}
\end{figure}

\subsection{Evaluating different training strategies.}

%Next, we choose the scattering model as the most accurate one and evaluate the proposed machine learning approach in its ability to substitute the physical model. 

\begin{table}[t]
\centering
\resizebox{\textwidth}{!}{
\begin{tabular}{ccccccccc}
\toprule 
Dataset &  & Individual ID &  & \multicolumn{2}{c}{Spectral MAE} &  & \multicolumn{2}{c}{Concentration MAE}\tabularnewline
\cmidrule{5-6} \cmidrule{6-6} \cmidrule{8-9} \cmidrule{9-9} 
 &  &  &  & Non-Scattering & Scattering &  & Strategy (a) & Strategy (b)\tabularnewline
\midrule 
 &  & 507 & & \num{0.012262} & \textbf{\num{0.008601}} & & \num{0.013588} & \textbf{\num{0.004318}}\tabularnewline
\cmidrule{2-9} \cmidrule{3-9} \cmidrule{4-9} \cmidrule{5-9} \cmidrule{6-9} \cmidrule{7-9} \cmidrule{8-9} \cmidrule{9-9} 
Broadband &  & 509 & & \num{0.010915} & \textbf{\num{0.009368}} & & \num{0.024570} & \textbf{\num{0.004812}}\tabularnewline
\cmidrule{2-9} \cmidrule{3-9} \cmidrule{4-9} \cmidrule{5-9} \cmidrule{6-9} \cmidrule{7-9} \cmidrule{8-9} \cmidrule{9-9} 
NIRS &  & 511 & & \num{0.008005} & \textbf{\num{0.006595}} & & \num{0.011586} & \textbf{\num{0.003408}}\tabularnewline
\cmidrule{2-9} \cmidrule{3-9} \cmidrule{4-9} \cmidrule{5-9} \cmidrule{6-9} \cmidrule{7-9} \cmidrule{8-9} \cmidrule{9-9} 
 &  & 512 & & \num{0.011992} & \textbf{\num{0.010882}} & & \num{0.012532} & \textbf{\num{0.004591}}\tabularnewline
\midrule 
 &  & 012-01 & & \num{0.032690} & \textbf{\num{0.024896}} & & \num{0.173464} & \textbf{\num{0.016395}}\tabularnewline
\cmidrule{2-9} \cmidrule{3-9} \cmidrule{4-9} \cmidrule{5-9} \cmidrule{6-9} \cmidrule{7-9} \cmidrule{8-9} \cmidrule{9-9} 
HELICoiD &  & 012-02 & &  \num{0.022411} & \textbf{\num{0.021926}} & & \num{0.149752} & \textbf{\num{0.025800}}\tabularnewline
\cmidrule{2-9} \cmidrule{3-9} \cmidrule{4-9} \cmidrule{5-9} \cmidrule{6-9} \cmidrule{7-9} \cmidrule{8-9} \cmidrule{9-9} 
 &  & 015-01 & & \num{0.063264} & \textbf{\num{0.025373}} & & \num{0.174669} & \textbf{\num{0.015381}}\tabularnewline
\bottomrule
\end{tabular}}
\\
\caption{Quantitative performance comparison of the different Beer-Lambert models and network training strategies on the test set of the two spectral datasets. To compare the two (non-scattering and scattering) models, we compute the mean absolute error of the spectral fit (denoted as 'Spectral MAE') between the ground truth observed and predicted signals. The two network training strategies are compared by assessing the mean absolute error of each strategy between the network and optimization-inferred concentrations (denoted as 'Concentration MAE') of all considered chromophores. In the case of the HELICoiD dataset, only pixels labeled as normal, tumor, or blood were considered for these computations. The best-performing model and strategy for each individual is highlighted in bold.}
\label{tab:mae_table}
\end{table}

Next, we evaluate the proposed machine learning approach in its ability to substitute both the linear absorption and the non-linear scattering model.

Fig. \ref{StrAB-NIRS} demonstrates the results of the experiment in which we test the network trained on synthetic data collected according to strategy (a) and on real data according to strategy (b), for both linear and non-linear models. For the linear case, both strategies are able to correctly infer the concentrations. The solution to the linear model can merely be found by a matrix multiplication, i.e. the pseudoinverse, which is why both strategies are able to very accurately predict the optimization-inferred concentrations.

For the non-linear case, strategy (b) provides qualitatively closer fits. We also tested this model for the Helicoid dataset, where we found highly matching results by the use of strategy (b), as seen in Fig. \ref{StrAB-Helicoid}.

These findings are also quantitatively supported by the results in Table \ref{tab:mae_table}. 
%The scattering model cannot be learned via a simple matrix computation due to the inherent non-linearity. 
%The proposed strategy (b) is able to capture non-linear physical effects beyond absorption properties, exclusively present when using strategy (a), such that almost indistinguishable predictions from the optimization-based results can be obtained.

\subsection{Computational time.}
\begin{SCfigure}
    \includegraphics[width=10cm]{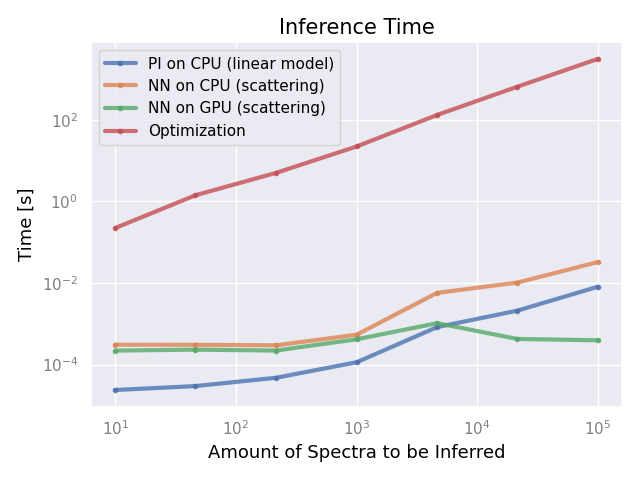}
  \caption{Comparison between inference time for various optimization approaches for varying number of spectra (from $10$ to $10^5$): including the pseudoinverse for the linear model (blue) and optimization-based (red) for the non-linear scattering model (both running on CPU), as well as network-based approach for scattering model running on CPU (orange) and GPU (green).}
\label{time}
\end{SCfigure}

Importantly, the proposed network-based optimization comes with a significant speed-up in computational time. In Fig. \ref{time}, we show a comparative analysis for performing chromophore composition inference using standard least-squares solvers (based on gradient update or pseudoinverse) and our proposed approach. The used spectra for this comparative analysis are taken from the broadband NIRS dataset assuming the scattering model, i.e. they are in the 740nm-900nm range, with a total of 244 measured wavelengths per spectrum, and the underlying chromophores are oxyhemoglobin, deoxyhemoglobin, and differential cytochrome-c-oxidase.

As solving the linear system using the pseudoinverse requires the least amount of matrix multiplication, this method provides the fastest compute. 
However, with the growth of the number of spectra for which we solve the optimization task, the matrix size for the inversion increases, and thus the computational time increases. Starting from ca. $10^4$ amount of spectra, the proposed network having a fixed amount of computational units becomes superior in terms of optimization time. Such runtime will remain approximately constant with a further increasing number of spectra, assuming sufficient GPU memory is available. More importantly, for non-linear systems, which are here represented as a Beer-Lambert model with the inclusion of scattering, one cannot utilize the pseudoinverse and has to resort to non-linear solvers like the ones based on gradient update. Such solvers are two to three orders of magnitude slower than the neural network approach, which has fixed compute time for linear and non-linear systems. As Fig. \ref{time} shows, it takes ca. 0.4 ms for the network to infer biochemical composition for $10^5$ spectra on NVIDIA GeForce MX450 with 2048 MiB. Overall, on our hardware, it takes between 2.5 and 3.1 seconds to run the neural network for one an image from the HELICoiD dataset (the largest among all tested data), from opening the normalized HSI image and loading the neural network into the GPU, to displaying the inferred concentrations.

\section{Discussion and Conclusion}

First, in this paper we wanted to address the limitation of the existing machine learning approaches to infer molecular composition using a physical model. Predominantly, the training within these approaches is performed on synthetic data produced by the physical model, analogously to the strategy (a) above. However, real spectroscopy measurements include a few factors not considered by the modeling, such as instrumentation and physiological noise or other non-linear optical phenomena. To close the gap between synthetic and real data, several works propose incorporating various kinds of noise into the synthetic data  \cite{scarbrough2024designing}. But the realism of the used noise formulations and their sufficiency to close the gap can still be questioned. Instead, the learning scheme proposed in this paper results in training and testing carried out on the same type of data obtained by processing real spectra with traditional least-squares optimisation. Thus, no noise model is required to achieve accurate predictions. 

However, we must admit that the predictions by the existing training approach are, nevertheless, close to the ground truth. Intuitively, this behavior is not within expectations as, again, the spectra of synthetic data on which the network was trained and real data on which it was evaluated notably vary. We attribute such behavior to the fact that the main spectral feature that networks learn to focus on is the global functional shape of the spectra. The local behavior of the spectral function is less informative, and thus, the local perturbations, i.e., random noise, do not affect the network performance significantly. 
This reasoning is further reinforced when comparing the network performance between bNIRS and HSI data. For the HSI data, the noise has a pronounced wavelength dependency, in contrast to bNIRS data (see Fig. \ref{helic_spect}), which results in the network performance (trained according to the existing approach)  degrading more notably compared to bNIRS data. 

Another explanation for this behavior can be that here, we try to learn with a neural network a solution to a linear system (or quasi-linear system, in the case of the model with scattering, where scattering contribution to the reflection spectra is minor compared to the absorption, as for bNIRS data). We hypothesize that upon network training on the dataset representing solutions to linear systems, the network weights are learned to minimize the contribution from non-linear network units (e.g., activation functions), as such units are unnecessary to establish a linear mapping. Generally speaking, the non-linear units are both a strength and a limitation of the networks. The strength is that they enable to learn an arbitrary mapping, while the limitation is the very sequential application of non-linear units that can cause the predictions to become highly unstable for input data unseen during training. But if the contribution from non-linear units of the network is minimized, we would not expect such instability of the network's predictions.

%Within expectation were, however, the results obtained by the strategy (b). In contrast to training on the synthetic data, strategy (b) allows us to infer molecular composition almost identical to the convex optimization solver for both bNIRS and HSI spectra. 

%Firstly, the Gaussian noise is propagated directly to the inferred concentrations. We can observe this in figure 6, as the Gaussian noise in the HbO2 signal decreases significantly going from strategy (a) to strategy (b). The same phenomenon is visible for the Helicoid dataset, where we observe that strategy (a) results in a grainy image. 

Second, we want to understand the limits of applicability of the linear model (with no scattering) for describing brain tissue spectroscopy. Note that under the assumption of negligible contribution from scattering in the modified Beer-Lambert law, the Eq. \ref{dBBL} results in a linear system. Such a system can be efficiently solved using the pseudoinverse, achieving close to real-time computation. However, in general, when including non-linear terms in the Beer-Lambert law formulation (scattering, non-linear absorption effects, etc.), the system of equations can not be solved anymore using pseudoinverse. Our analysis reveals that including scattering can often be necessary to describe the spectroscopy measurements, especially for HSI data. But also for the bNIRS data, we have qualitatively observed that the nonlinear model seems to fit the wavelength range between 740nm and 780nm particularly well, Fig. \ref{relative}. This finding was within expectations since the range between 780nm and 900nm is not regarded to be dominated by changes in scattering and has therefore been widely used in broadband NIRS technologies with concentrations inferred through the Beer-Lambert law solely considering absorption \cite{kaynezhad2019quantification}. 

Importantly, as a by-product of our analysis, when we test the spectral unmixing on the brain tissue HSI dataset, we observe that the molecular inference driven by the physical model leads to better contrasting the blood vessel tree compared to the RGB image, Fig. \ref{helic_res} (middle), and capturing metabolic activity (right). Image semantics related to both hemodynamic and metabolic tissue properties could be highly valuable in assisting neurosurgeons during the process of tumor removal. Knowledge of the topology of the blood vessel tree would minimize the amount of undesirable resection of the vessels leading to bleeding. In turn, understanding the metabolic activity across the 2D field of view would allow for better separation of pathological tissue from the healthy parenchyma, reducing the time spent on intra- and post-operative biopsy.  

To conclude, in the paper, we present a data-driven concept for inferring molecular composition change from diffuse spectroscopy of brain tissue. We test the approach on various datasets (bNIRS and HSI) and physical models of different complexity (with and without scattering, i.e., linear and non-linear systems). Importantly, we evaluate different training strategies for neural-networks-based molecular prediction. The proposed strategy provides predictions that are nearly identical to the traditional least-squares-fit method, making the learnable solver an accurate alternative. 
Finally, the method achieves subsecond time for simultaneous inference of molecular composition across a large number of spectra, allowing for real-time tissue characterization using bNIRS and HSI imaging modalities.

\subsection*{Disclosures}
The authors declare no financial conflict of interest.

\subsection* {Code, Data, and Materials Availability} 
In support of open science, all our code is available on github under  \url{https://github.com/HyperProbe/SpectraFit}. The HSI data is publicly available by the HELICoiD project. The bNIRS data may be obtained upon request. Both datasets received ethical approval as described in the corresponding papers \cite{kaynezhad2019quantification,fabelo2019vivo}.

\subsection* {Acknowledgments}
The HyperProbe consortium has received funding from the European Union’s Horizon Europe research under grant agreement No. 101071040. UCL is supported by UKRI grant No. 10048387.

%%%%% References %%%%%

\bibliography{report}   % bibliography data in report.bib
\bibliographystyle{spiejour}   % makes bibtex use spiejour.bst

\listoffigures
\listoftables

\newpage
\section{Appendix}

\subsection{Absorption Spectra of Chromophores}

\begin{figure}[h]
	\begin{centering}
		\includegraphics[width=\textwidth]{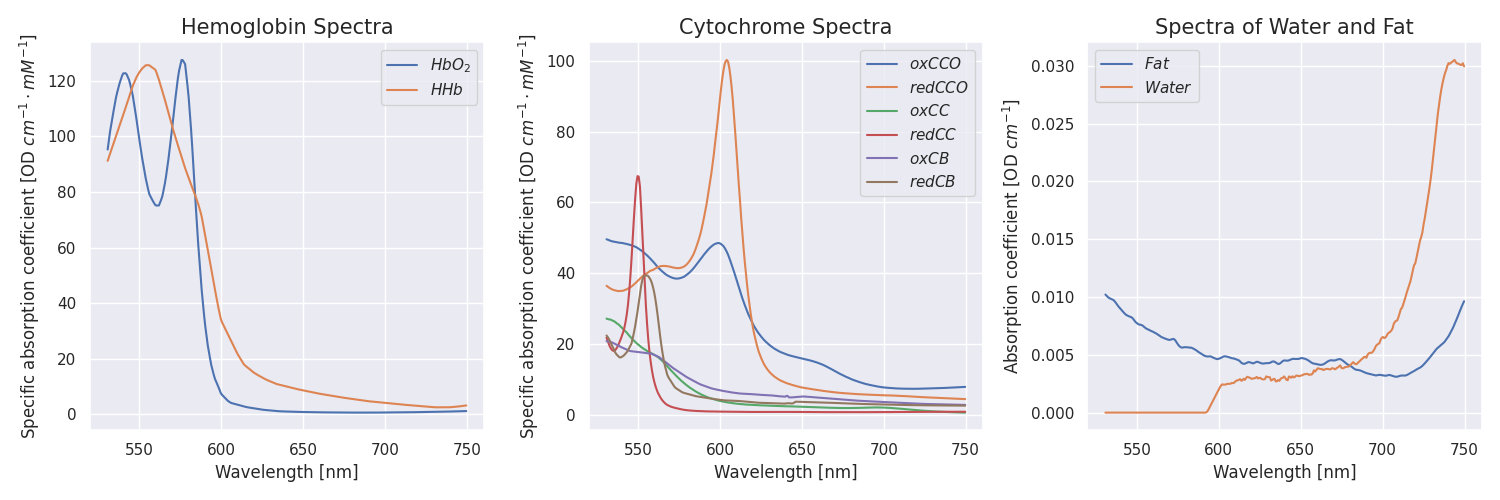}
		\par\end{centering}
	\caption{Absorption coefficients of the fitted chromophores \cite{UCL-NIR,Matcher1994, Zijlstra2021, Heinrich1981} used for the Helicoid dataset. The first plot shows absorption coefficients of oxyhaemoglobin and deoxyhaemoglobin. The second plot shows absorption coefficients of cytochrome-c-oxidase, cytochrome-c, and cytochrome-b in oxidized and reduced form, respectively. Units of these absorption coefficients are per cm per millimole, as they represent concentrations. The third plot shows absorption coefficients of fat and water in the form of volumetric content, with unit per cm.}
	\label{fig:spectra-helicoid}
\end{figure}

For the broadband NIRS dataset, we used oxyhemoglobin, deoxyhemoglobin, and differential cytochrome-c-oxidase as absorbing chromophores. These have been used extensively in literature in the context of NIRS imaging \cite{kaynezhad2019quantification}. For HSI brain tumor imaging, there is no standardized chromophore set in literature that could be used for fitting observed attenuations. We therefore resorted to the assumptions explained in the main text, and report absorption spectra of these fitted chromophores in  Fig. \ref{fig:spectra-helicoid}.

\subsection{Dataset Generation Details}
To generate the synthetic datasets necessary to train the neural networks with strategy (a), we uniformly sample parameters from certain ranges. For the broadband NIRS dataset, the selected ranges are shown in Table \ref{tab:generation_ranges}. We also report the physiologically used ranges for the scattering parameters $a_1,b_1$, used in both datasets, and the corresponding reference upon which they were based. For the Helicoid dataset, we do not make any direct physiological assumptions as we instead use minima and maxima of the parameters found during optimization.
\begin{table}[H]
    \centering
    \begin{tabular}{ccccc}
    \toprule 
    Parameter Type & Parameter & Minimum & Maximum & Reference\tabularnewline
    \midrule
    \midrule 
    \multirow{3}{*}{\makecell{Concentration \\ (Broadband NIRS)}} & $HbO_{2} \text{  } [\frac{mM}{cm}]$ & $-0.5$ & $0.5$ & \cite{kaynezhad2019quantification} \tabularnewline
    \cmidrule{2-5} \cmidrule{3-5} \cmidrule{4-5} \cmidrule{5-5} 
     & $HHb \text{  } [\frac{mM}{cm}]$ & $-0.5$ & $0.5$ & \cite{kaynezhad2019quantification} \tabularnewline
    \cmidrule{2-5} \cmidrule{3-5} \cmidrule{4-5} \cmidrule{5-5} 
     & diffCCO $[\frac{mM}{cm}]$ & $-0.25$ & $0.25$ & \cite{kaynezhad2019quantification} \tabularnewline
    \midrule 
    \multirow{2}{*}{Scattering} & $a_{1} \text{  } [\frac{1}{cm}]$ & $0$ & $100$ & \cite{jacques2013optical}\tabularnewline
    \cmidrule{2-5} \cmidrule{3-5} \cmidrule{4-5} \cmidrule{5-5} 
     & $b_{1} \text{  } [1]$ & $0$ & $5$ & \cite{jacques2013optical} \tabularnewline
    \bottomrule
    \end{tabular}
    \caption{Ranges from parameters whose range was chosen based on physiological assumption, with corresponding reference. }
    \label{tab:generation_ranges}
\end{table}

\subsection{Neural Network Training}

All networks were trained with the Adam optimizer. The Ray Tune framework \cite{liaw2018tune} automatically finds neural network hyperparameters that would otherwise be difficult to manually tune and find. It selected the optimal number of hidden layers $H$, the network width $W$, the learning rate $\lambda$, the activation function $f$, and the batch size $B$ in 200 trials of random search from the following ranges: $H \in \{0,1,2,3\}$ or $H \in \{0,1,2,3,4\}$ (depending on whether training was being performed on the broadband NIRS or HELICoiD dataset), $W \in [1,64]$, $\lambda \in [10^{-4}, 10^{-1}]$, $f \in \{ \text{ELU, Hardshrink, LeakyReLU}\}$, and $B \in \{ 32, 64, 128, 256\}$ or $B \in \{ 32, 64, 128, 256, 512, 1024, 2048 \}$ (depending on the dataset). Slightly different ranges were used for the HELICoiD dataset to account for a possibly larger needed computational complexity, due to the inherently more complex dataset. The found optimal parameters are reported in Table \ref{tab:neural_network_appendix}.
\begin{table}[H]
    \centering
    \begin{tabular}{ccccccc}
    \toprule 
    Dataset & Neural & Hidden & Network & Learning & Activation & Batch\tabularnewline
    \& Model & Network & Layers & Width & Rate & Function & Size\tabularnewline
    \midrule
    Broadband NIRS & Strategy (a) & 2 & 35 & $5.28832 \cdot 10^{-3}$ & Leaky ReLU & 128\tabularnewline
    \cmidrule{2-7} \cmidrule{3-7} \cmidrule{4-7} \cmidrule{5-7} \cmidrule{6-7} \cmidrule{7-7} 
    (Linear) & Strategy (b) & 2 & 16 & $1.7495724 \cdot 10^{-3}$ & Leaky ReLU & 32\tabularnewline
    \midrule 
Broadband NIRS & Strategy (a) & 1 & 62 & $7.86158\cdot10^{-4}$ & Leaky ReLU & 32\tabularnewline
    \cmidrule{2-7} \cmidrule{3-7} \cmidrule{4-7} \cmidrule{5-7} \cmidrule{6-7} \cmidrule{7-7} 
    (Non-Linear) & Strategy (b) & 2 & 53 & $5.94006\cdot10^{-4}$ & ELU & 32\tabularnewline
    \midrule
    HELICoiD & Strategy (a) & $1$ & $31$ & $6.64275\cdot10^{-3}$ & ELU & 512\tabularnewline
    \cmidrule{2-7} \cmidrule{3-7} \cmidrule{4-7} \cmidrule{5-7} \cmidrule{6-7} \cmidrule{7-7} 
    (Non-Linear) & Strategy (b) & $2$ & $46$ & $3.52619\cdot10^{-4}$ & ELU & 32\tabularnewline
    \bottomrule
    \end{tabular}
    \caption{Parameters listed in the text, found by the RayTune library for each dataset (Broadband NIRS and HELICoiD), model (linear absorption and non-linear scattering), and neural network training strategies (a) and (b), respectively.}
    \label{tab:neural_network_appendix}
\end{table}

\end{spacing}
\end{document}